\documentclass[pra,aps,twocolumn,superscriptaddress,floatfix,showpacs,longbibliography]{revtex4-2}
\usepackage{graphicx}
\usepackage{dcolumn}
\usepackage{bm}
\usepackage{upgreek}
\usepackage{natbib}
\usepackage[normalem]{ulem}
\usepackage{amsmath,amssymb}
\usepackage{dsfont}
\usepackage[english]{babel}
\usepackage{color}

\begin{document}
\title{$\mathbb{Z}_n$ symmetry broken supersolid in spin-orbit-coupled Bose-Einstein condensates}
\author{Ze-Hong Guo}
\affiliation{Guangdong Provincial Key Laboratory of Quantum Engineering and
Quantum Materials, School of Physics and Telecommunication Engineering, South
China Normal University, Guangzhou 510006, China}
\affiliation{Guangdong-Hong Kong Joint Laboratory of Quantum Matter, Frontier
Research Institute for Physics, South China Normal University, Guangzhou
510006, China}
\author{Qizhong Zhu}
\email{qzzhu@m.scnu.edu.cn}
\affiliation{Guangdong Provincial Key Laboratory of Quantum Engineering and
Quantum Materials, School of Physics and Telecommunication Engineering, South
China Normal University, Guangzhou 510006, China}
\affiliation{Guangdong-Hong Kong Joint Laboratory of Quantum Matter, Frontier
Research Institute for Physics, South China Normal University, Guangzhou
510006, China}

\date{\today}

\begin{abstract}
  Supersolid is an exotic state of matter characterized by both superfluid properties and periodic
 particle density modulation, due to spontaneous breaking of U(1) gauge symmetry and spatial translation symmetry, respectively. For conventional supersolids,
continuous translation symmetry breaking is accompanied by one gapless Goldstone mode in the excitation spectra.
 An interesting question naturally arises: what is the consequence of breaking discrete translation symmetry for supersolids?
  In this work, we propose the concept of $\mathbb{Z}_n$ supersolid resulting from spontaneous breaking of
 a discrete $\mathbb{Z}_n$ symmetry, or equivalently, a discrete translation symmetry. This
 $\mathbb{Z}_n$ supersolid is realized in the stripe phase of spin-orbit-coupled Bose-Einstein condensate under an external periodic potential with period $1/n$ of intrinsic stripe period. For $n\geq2$, there are $n$ degenerate ground states with spontaneously 
broken lattice translation symmetry.
The low-energy excitations of $\mathbb{Z}_n$ supersolid include a pseudo-Goldstone mode, whose excitation gap at long wavelength limit is found to decrease fast with $n$.
We further numerically show that, when confined in a harmonic trap, a spin-dependent perturbation can result in the transition between degenerate ground states of $\mathbb{Z}_n$ supersolid. With the integer $n$ tunable using the experimental technique of generating subwavelength optical lattice, the $\mathbb{Z}_n$ supersolid proposed here offers a cold atom platform to simulate physics related with generic $\mathbb{Z}_n$ symmetry breaking,
which is interesting not only in the field of cold atoms, but also in particle physics and cosmology.   
\end{abstract}

\maketitle

\section{introduction}

Supersolid as a fascinating state of matter has received much attention in the field of cold atoms over the last few years.
It features superfluid properties and at the same time crystalline order in particle density \cite{boninsegni_TextitColloquium_2012},
from spontaneous breaking of U(1) gauge symmetry and continuous spatial translation symmetry, respectively.
Several mechanisms in cold atoms have been proposed to successfully realize a supersolid in experiments, including the spin-orbit-coupled Bose-Einstein condensate (BEC) \cite{wang_SpinOrbit_2010,ho_BoseEinstein_2011,li_Stripe_2017,chen_Quantum_2018,baena_Supersolid_2020},
the BEC coupled with optical cavities \cite{leonard_Supersolid_2017} and dipolar gases with roton excitation spectrum \cite{tanzi_Observation_2019,bottcher_Transient_2019,chomaz_LongLived_2019,
guo_Lowenergy_2019,tanzi_Supersolid_2019,norcia_Twodimensional_2021,petter_Bragg_2021}. 
Supersolids resulting from spontaneous breaking of continuous translation symmetry all share the feature of one gapless Goldstone mode
in the excitation spectrum associated with this symmetry breaking \cite{li_Superstripes_2013,xia_Metastable_2023}, in stark contrast to superfluids with ``forced'' density modulation by external potential.

Under external periodic potential, the spatial translation symmetry of this system is explicitly broken. The interplay between spin-orbit coupling and
external periodic potential has been shown to result in rich interesting physics, such as flat band \cite{zhang_BoseEinstein_2013}, supersolid with
topologically nontrivial spin textures \cite{han_Supersolid_2015}, incommensurate stripe phases \cite{chen_Groundstate_2016}, crossover of commensurate-incommensurate supersolid \cite{li_Commensurateincommensurate_2020}, unconventional excitations of stripes \cite{maeland_Plane_2020} and non-uniform
superfluid-insulator transitions \cite{yamamoto_Quantum_2017,yamamoto_Supersolid_2022}, etc.
 The periodic potential has been experimentally demonstrated to be favourable for the formation of
 stripe density modulation \cite{bersano_Experimental_2019}. Much more recently, the effect of periodic potential on the excitation spectrum of stripe phase is investigated by Li {\it et al.} \cite{li_PseudoGoldstone_2021},
and a gapped pseudo-Goldstone mode is found in distinction to the gapless Goldstone mode in the case of vanishing periodic potential.
Those authors have considered the scenario that the period of external potential just coincides with the intrinsic period of stripe, and hence no other
symmetry is broken in the stripe phase than the U(1) gauge symmetry. The question naturally arises as to what if the external potential has a different period? In particular, if the external periodic potential has period $1/n$ of intrinsic stripe period, how will the supersolid resulting from a discrete symmetry breaking be different with one resulting from continuous symmetry breaking?

In this work, we aim to answer these interesting questions above by considering a spin-orbit-coupled BEC with external periodic potential, whose period is $1/n$ of intrinsic stripe period.
We find that the stripe phase at ground state is associated with a discrete $\mathbb{Z}_n$ symmetry breaking,
in contrast to the vanishing periodic potential case where a continuous translation symmetry is broken.
As will be shown below, the spatial translation of stripe density can be uniquely mapped to a relative phase degree of freedom in stripe wave function,
and the $\mathbb{Z}_n$ symmetry of the relative phase corresponds to a discrete translation symmetry of stripe density. The $\mathbb{Z}_n$ symmetry breaking here
means the ground state is only invariant under translation by the intrinsic stripe period instead of the external potential period and there exist
multiple energetically degenerate ground states each differing by the translation of external potential period.
This phase resembles the theoretically studied supersolid in lattice models without spin-orbit coupling \cite{boninsegni_Supersolid_2005,scarola_Quantum_2005,heidarian_Persistent_2005,melko_Supersolid_2005,scarola_Searching_2006,danshita_Stability_2009,
trefzger_PairSupersolid_2009,pollet_Supersolid_2010,zhang_Supersolid_2011,ohgoe_Commensurate_2012,li_Latticesupersolid_2013,zhang_Quantum_2016,masella_Supersolid_2019},
where the lattice supersolid breaks lattice translation symmetry with a longer period.
Those phases were proposed based on finite-range extended Bose-Hubbard models, and till now have yet to be experimentally observed. In contrast, here the atoms we consider interact with each other through $s$-wave contact interaction,
and the model is more feasible in experiments based on the current experimental setup.
We note that $\mathbb{Z}_n$ symmetry breaking in similar systems was also pointed out in previous work \cite{yamamoto_Quantum_2017,yamamoto_Supersolid_2022}, but the focus was on the ground state phase diagram
and the physical consequence of $\mathbb{Z}_n$ symmetry breaking in supersolid, such as low-energy excitations, has not been explored.
In this paper, we study in detail the consequence associated with this $\mathbb{Z}_n$ symmetry breaking, including the degenerate ground states, 
the low-energy excitations of $\mathbb{Z}_n$ supersolid, and its dynamical feature in experiments.
We find the fast decay of excitation gap in pseudo-Goldstone mode with $n$, and unique dynamical behaviour of $\mathbb{Z}_n$ supersolid in the presence of harmonic trap, which can serve as an experimental signature for its observation.
This model thus offers an experimentally feasible platform to study the physical consequence associated with generic $\mathbb{Z}_n$ symmetry breaking,
which is not only interesting in the research of supersolid, but also has implication for similar phenomena resulting from $\mathbb{Z}_n$ symmetry
breaking in the fields of particle physics and cosmology \cite{sikivie_Axions_1982,freese_Inflating_2005,sierra_Dynamical_2015,cherman_Lifetimes_2021,wu_Classification_2022,wu_Collapsing_2022}.

The rest of paper is laid out as follows. 
First, in Sec. \ref{sec2}, we start from the single-particle model of spin-orbit-coupled bosons under external periodic potential, treat the
atomic interaction within the mean-field theory, and demonstrate the existence of degenerate ground states. 
Then, in Sec. \ref{sec3}, we calculate the excitation spectrum of stripe phase with Bogoliubov theory, reveal the dependence of excitation gap
on $n$, and explain the behaviour of gap dependence in terms of the effective Hamiltonian.
In Sec. \ref{sec4}, we show the distinct dynamical properties of stripe above ground state under spin-dependent perturbation for different $n$, 
which can serve as the experimental signature of $\mathbb{Z}_n$ symmetry broken supersolid, and 
demonstrate the transition among different degenerate ground states during temporal evolution. 
Finally, we summarize and conclude in Sec. \ref{sec5}.

\section{$\mathbb{Z}_n$ symmetry breaking and degenerate ground states}
\label{sec2}

\begin{figure}[tbp] \centering
    \includegraphics[width=0.99\linewidth]{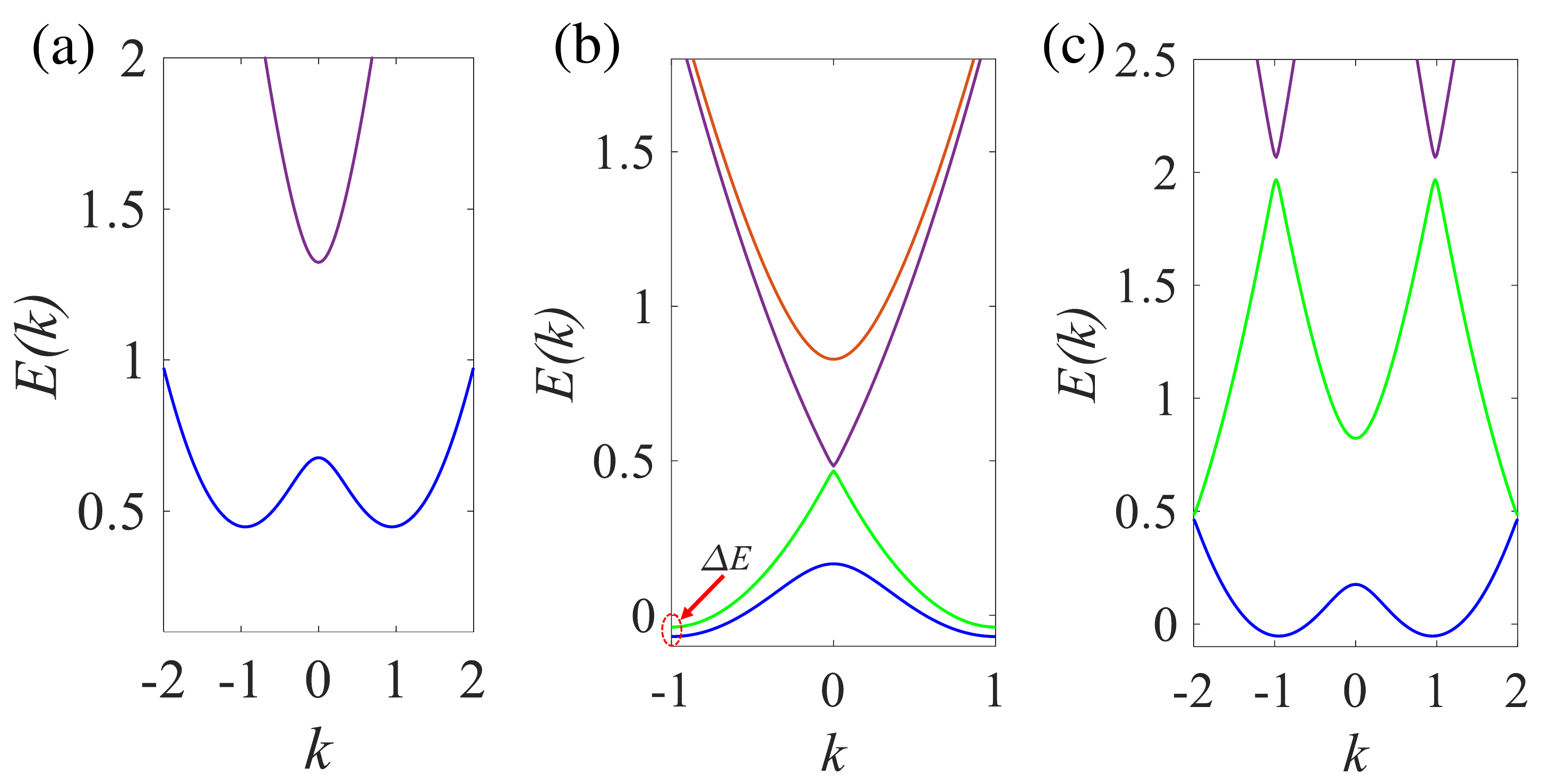}
    \caption{(a) Single-particle dispersion of spin-orbit-coupled bosons in free space. (b) and (c) Non-interacting Bloch band of Eq. \ref{eq1} for $n=1$ and $n=2$, respectively. The energy gap $\Delta E$ at dispersion minima in (b) is also indicated by red arrow, and is further studied in Fig. \ref{fig5}. The other parameters are $V_{0}=0$, $\Omega=1.294$ for (a), and $V_{0}=0.2$, $\Omega=1.294$ for (b) and (c). }
    \label{fig1}
  \end{figure}

We consider a one-dimensional spin-orbit-coupled bosonic system, where the spin-orbit coupling is engineered by two Raman laser beams which
introduce momentum transfer in $x$ direction to couple two pseudospin states of atoms \cite{li_Stripe_2017}. An external periodic potential generated by
subwavelength optical lattice is also imposed, 
and thus this system can be described by the following single-particle Hamiltonian
\begin{equation}
  \label{eq1}
  \hat{H}_{0}=\frac{\hbar^{2}}{2m}\left(-i\partial_x-k_{0}\sigma_{z}\right)^{2}+\frac{\Omega}{2}\sigma_{x}+V(x),
  \end{equation}
  where $k_{0}$ is the strength of spin-orbit coupling, $\Omega$ is the Raman coupling strength, 
  $\sigma_{i}$ ($i=x,z$) is the Pauli matrix in pseudospin space, and $V(x)$ is the external periodic potential.
  Throughout this paper, we have set $k_{0}$ and recoil energy $E_r=\hbar^{2}k_{0}^{2}/2m$ as the units of wavenumber and energy, respectively. In the absence of external potential, 
  the dispersion of single-particle Hamiltonian for small $\Omega$ exhibits a double-well structure as shown in Fig. \ref{fig1}(a), with $\pm k_s=\pm k_0\sqrt{1-\Omega^2/16E_r^2}$ denoting the wavenumber at dispersion minima \cite{li_Quantum_2012}. 
  The modification of $k_s$ at ground state brought by weak atomic interaction is very slight \cite{li_Quantum_2012,xia_Metastable_2023} and is neglected in the present work.
   Previous studies
  have focused on the case that the imposed optical lattice potential has the same period with that of stripe density $d_s=\pi/k_s$
  \cite{li_PseudoGoldstone_2021}, which uniquely pins the density modulation at ground state. 
  Here, we consider an optical lattice potential $V(x)=V_{0}\cos{(2nk_{s}x)}$, with a general integer $n=1,2,3\dots$. In other words, the
  lattice period is $1/n$ of intrinsic stripe period $d_s$ and the previous work corresponds
  to the case of $n=1$ \cite{li_PseudoGoldstone_2021}. 
  The integer $n$ is tunable by adjusting the period of optical lattice potential, e.g., using the experimental technique
  of creating subwavelength optical lattice as successfully demonstrated in Ref. \cite{anderson_Realization_2020}. 
  This external potential introduces direct coupling between plane waves with momenta differing by $2nk_{s}$,
  and there also exists coupling between momenta differing by $2k_{s}$ due to the existence of intrinsic stripe density modulation and atomic interaction, which can be understood as an effective periodic
  potential within the mean-field theory.
Note that the period of stripe density is still $d_s$.

  The Bloch band dispersions for $n=1$ and $n\neq1$
  are shown in Figs. \ref{fig1}(b) and (c), respectively. It is clear that 
  for $n=1$, there exists an energy gap $\Delta E$ at dispersion minima, coinciding with the boundary
  of the first Brillouin zone (BZ). This energy gap is induced by direct coupling
  between states with momenta differing by $2k_s$, due to the presence of external periodic potential. 
  In contrast, for $n\geq2$, the gap induced by periodic potential appears at BZ boundary instead of dispersion minima.

In the stripe phase, the spin-orbit-coupled BEC interacting through $s$-wave contact interaction can be well treated
within the mean-field theory \cite{li_Quantum_2012}, where the energy functional reads,
\begin{align}
  {E}\left(\Psi_{\uparrow},\Psi_{\downarrow}\right) = & \int\mathrm{d}x\left(\Psi_{\uparrow}^{*},\Psi_{\downarrow}^{*}\right)\hat{H}_{0}\left(\begin{array}{c}
  \Psi_{\uparrow}\\
  \Psi_{\downarrow}
  \end{array}\right)
  +g_{\uparrow\downarrow}|\Psi_{\uparrow}|^{2}|\Psi_{\downarrow}|^{2}\nonumber\\
   & +\frac{g}{2}\left(|\Psi_{\uparrow}|^{4}+|\Psi_{\downarrow}|^{4}\right).
   \label{ek}
  \end{align}
  Here $g$ and $g_{\uparrow\downarrow}$ denote the interaction strengths between atoms in the same and different pseudospin states, respectively,
  which are related with the $s$-wave scattering lengths between same and different atomic
  hyperfine states. We have assumed that the two intra-spin interaction strengths are the same, with $g_{\uparrow\uparrow}=g_{\downarrow\downarrow}=g$.
Through minimization of above energy functional,
one can obtain the stationary Gross-Pitaevskii equation (GPE) 
which holds for the stripe ground state, 
\begin{equation}
  \label{muk}
  \mu\Psi=\left[\hat{H}_{0}+\left(\begin{matrix}
  U_{\uparrow} & 0 \\ 
  0 & U_{\downarrow}
  \end{matrix}\right)\right]\Psi,
\end{equation}
with interaction terms $U_{\uparrow}=g|\Psi_\uparrow|^2+g_{\uparrow\downarrow}|\Psi_\downarrow|^2$ and $U_{\downarrow}=g|\Psi_\downarrow|^2+g_{\uparrow\downarrow}|\Psi_\uparrow|^2$. $\Psi=\left(\Psi_\uparrow,\Psi_\downarrow\right)^{T} $ is the two-component spinor wave function and $\mu$
  is the chemical potential. 
We use the following stripe wave function ansatz to characterize the stripe state,
\begin{equation}
  \label{eq3}
  \begin{pmatrix}
  \Psi_\uparrow \\
  \Psi_\downarrow  
  \end{pmatrix}=\sqrt{\rho_0}\sum_{l=-2L-1}^{2L+1}\begin{pmatrix}
  a_l \\
  b_l
  \end{pmatrix}e^{i l k_s x},
  \end{equation}
  where $L$ is a positive integer denoting the cutoff of momentum summation and the spacing of index $l$ is $2$. 
  The optimal wavenumber in stripe wave function at ground state has been taken to be $k_s$, since the interaction induced modification to $k_s$ is very small \cite{xia_Metastable_2023}.
   $\rho_0$ is the average condensate density.
  The expansion coefficients satisfy the relation $|b_l|=|a_{-l}|$ as the system is invariant under the combined operation of spatial inversion and spin flip.

  \begin{figure}[tbp] \centering
    \includegraphics[width=0.99\linewidth]{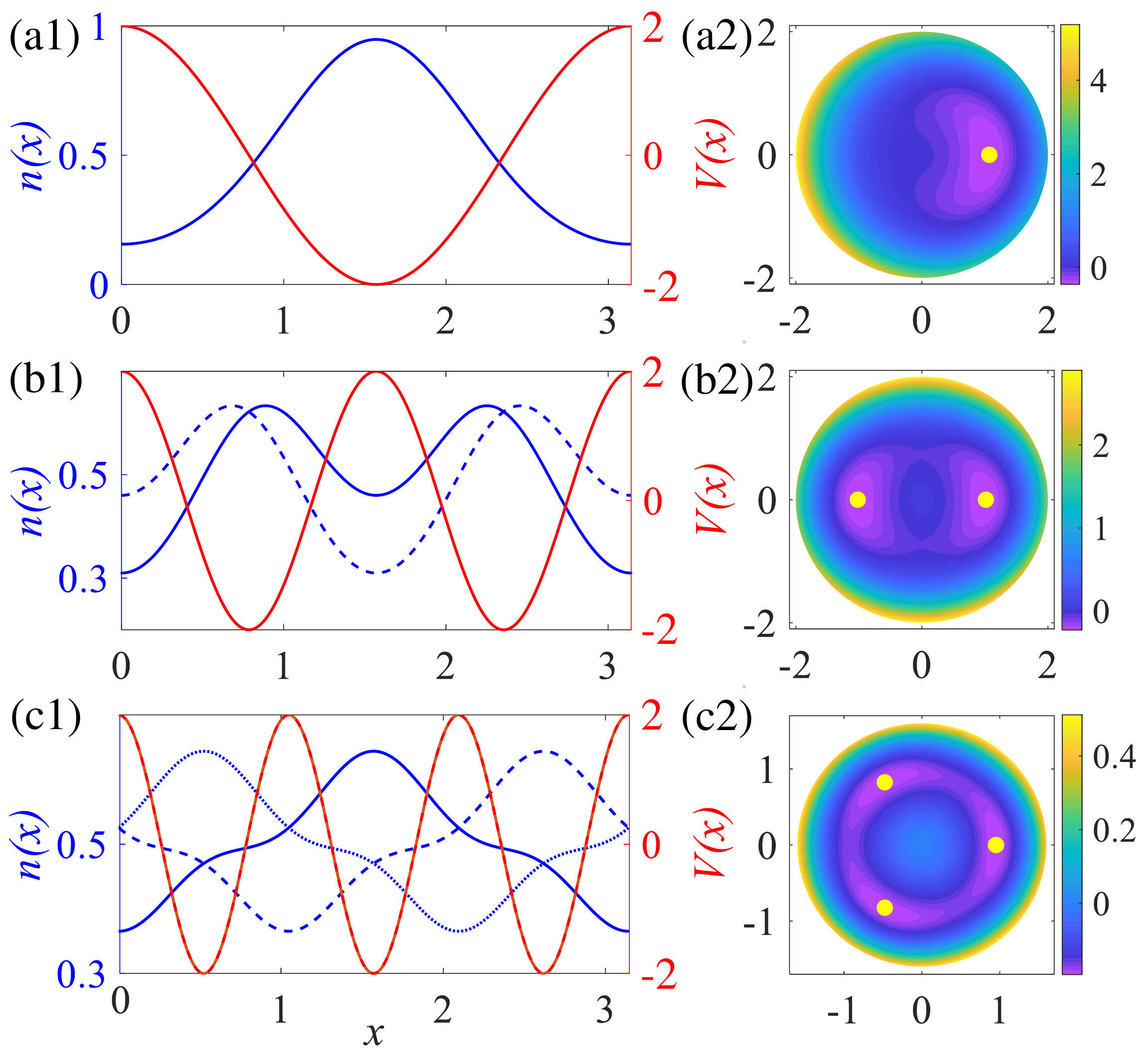}
    \caption{(a1)-(c1) Stripe density in the cases of $n=1$, $n=2$, and $n=3$, respectively. 
    Blue lines with different line types denote the density distribution (left vertical axis) of different ground states and red lines represent the external potential as a reference (right vertical axis).  
    (a2)-(c2) Effective potential energy $V_\mathrm{eff}(\sqrt{\rho_0},\theta)$ in the polar plane used to illustrate the discrete symmetry breaking corresponding to (a1)-(c1). 
    Yellow dots denote the discrete values of $\theta$ at potential minima, corresponding to different ground states.
    The other parameters are $\rho_0g=0.968$, $\rho_0 g_{\uparrow\downarrow}=0.88$, $\Omega=1.294$, and $V_{0}=2$. \label{fig2}}
  \end{figure}

  Plugging the above ansatz into Eq. (\ref{muk}) and solving a set of nonlinear equations 
  which the expansion coefficients $a_l$ and $b_l$ satisfy, we can obtain the ground state wave function, the corresponding density distribution $n(x)=|\Psi_\uparrow|^2+|\Psi_\downarrow|^2$, as well as the chemical potential $\mu$.
In realistic calculations, we have chosen $L=3$ ($l=-7, -5, -3, -1, 1, 3, 5, 7$), which gives results accurate enough as confirmed by both our calculation with larger $L$ and other work \cite{martone_Supersolid_2021}.  
  Note that there are two phase degrees of freedom in the coefficients, where the overall phase corresponds to the
  U(1) gauge symmetry breaking associated with superfluidity, and the relative phase corresponds to spatial translation of stripe density.
  The U(1) gauge symmetry is due to the total atom number conservation, which is obviously satisfied by the Hamiltonian considered here irrespective of the
  form of external potential,
  and the focus of this paper is on another symmetry breaking associated with the relative phase degree of freedom, whose character can be conveniently manipulated
  with the form of external potential.
To illustrate this point, we start from
  the energy functional and inspect whether it is invariant for certain choices of relative phase. If so,
  there exist multiple degenerate ground states and symmetry breaking associated with relative phase will occur. 
  In the absence of external periodic potential, it is obvious that, the following states with the wave function
     \begin{equation}
  \label{eq5}
  \begin{pmatrix}
  \Psi_\uparrow \\
  \Psi_\downarrow  
  \end{pmatrix}=\sqrt{\rho_0}\sum_{l=-2L-1}^{2L+1}\begin{pmatrix}
  a_l \\
  b_l
  \end{pmatrix}e^{i l (k_s x+\theta)},
  \end{equation} 
all have the same energy for arbitrary relative phase $\theta$ between positive and negative momenta. Here a transform is introduced in the
coefficients, i.e., $a_l\rightarrow a_l\exp(il\theta)$, $b_l\rightarrow b_l\exp(il\theta)$,
corresponding to
  spatial translation of stripe density by $\theta/k_s$, and the system energy remains unchanged for $V(x)=0$. Nevertheless, in the presence of periodic potential, the original Hamiltonian now only has discrete spatial translation symmetry,
 and the symmetry reduction term, i.e., the potential energy reads,
   \begin{align}
    E_{v}(\theta) = & \int_0^{d_s} \mathrm{d}x  n(x)V(x) \nonumber\\
    = & \frac{\rho_0 V_{0}}{2}\sum_{l_{1}-l_{2}=\pm2n}^{}(a_{l_{1}}^*a_{l_{2}}+b_{l_{1}}^*b_{l_{2}})e^{i(l_{2}-l_{1})\theta}.
       \label{ek_v}
  \end{align}
 This form results from the fact that the periodic potential $V_{0}\cos(2nk_{s}x)$ only
 couples plane waves with momenta differing by $\pm2nk_s$. 
  For the periodic potential chosen here, the coefficients $a_l$ and $b_l$ at ground state can all be taken to be real, along with
  $\theta=0$ in Eq. \ref{eq5}. In addition, with the relation
  $b_l=a_{-l}$, $E_{v}$ can thus be simplified as $E_{v}(\theta) = \rho_0 V_{0}\sum_{l_{1}-l_{2}=2n}^{}a_{l_{1}}a_{l_{2}}\cos(2n\theta)$.
  Clearly, the potential energy minimum is unique for $n=1$ and for $n\geq2$ resides at a series of discrete $\theta$, i.e., at 
  $\theta_m=m\pi/n$, $m=0,1,\dots, n-1$. These discrete $\theta_m$ in the transform of the coefficients $a_l$ and $b_l$ constitute the $\mathbb{Z}_n$
  discrete group, which is a subgroup of ``U(1)'' group in terms of continuous $\theta$. Since $\pi/nk_s=d_s/n$,
  these ground states with discrete $\theta_m$ are related with each other by spatial translation of stripe density in multiple of potential period.
  So there exist $n$ degenerate ground states with density modulation differing by spatial transition in multiple of $d_s/n$.
  The density of ground states for different $n$ is shown in Fig. \ref{fig2}. 
  As shown in Figs. \ref{fig2}(a1)-(c1), in the case of $n\neq1$, there exist $n$ degenerate ground states, and each can be obtained from another by spatial translation of lattice period.

The method of effective potential is often used to illustrate the spontaneous symmetry breaking, with the celebrated example being the ``Mexican hat'' profile
in U(1) gauge symmetry breaking.
In analogy, we show in the right panel of Fig. \ref{fig2} the profile of effective potential energy as a function of $\sqrt{\rho_0}$, $\theta$ in the polar plane, defined as 
$V_\mathrm{eff}(\sqrt{\rho_0},\theta)\equiv E_v(\rho_0,\theta)-\mu \rho_0+E_\mathrm{int}$, with the interaction energy $E_\mathrm{int}=\int\mathrm{d}x\,\left\{{g}/{2}\left(|\Psi_{\uparrow}|^{4}+|\Psi_{\downarrow}|^{4}\right)+g_{\uparrow\downarrow}|\Psi_{\uparrow}|^{2}|\Psi_{\downarrow}|^{2}\right\}$.
In the absence of external potential, $V_\mathrm{eff}(\sqrt{\rho_0})$ has the feature of ``Mexican hat'' with potential minima independent of $\theta$, which is a common feature for systems with continuous symmetry.
With the external periodic potential, the profile of $V_\mathrm{eff}(\sqrt{\rho_0},\theta)$ is depicted in Figs. \ref{fig2}(a2)-(c2).
Clearly, now the ``Mexican hat'' is deformed with potential minima located at $n$
discrete values of $\theta$, a feature of generic $\mathbb{Z}_n$ invariant potential (see for example, Ref. \cite{wu_Classification_2022}), which just correspond to $n$ degenerate ground states differing by spatial translation of lattice period.
Note also that the potential barrier between degenerate minima is dramatically reduced with the increase of $n$, showing the trend of approaching
``Mexican hat'' potential in the large $n$ limit.

\section{Elementary excitations of $\mathbb{Z}_n$ supersolid}
\label{sec3}

With external periodic potential, the continuous spatial translation symmetry is explicitly broken, and thus the gapless Goldstone mode
of a conventional supersolid now becomes gapped. In the view of effective potential, the degenerate potential minima are now discrete and
separated by finite potential barrier, such that any deviation from the minima will cost energy.
This leads to the interesting concept of pseudo-Goldstone mode, and has been studied
recently in the case of $n=1$ \cite{li_PseudoGoldstone_2021}. Here we focus on the case of $n\geq2$ and explore the properties of pseudo-Goldstone mode resulting from a $\mathbb{Z}_n$ symmetry breaking. From heuristic argument, for $n\rightarrow\infty$,
$\mathbb{Z}_n$ group will asymptotically approach a continuous group, i.e., ``U(1)'' group in terms of the relative phase $\theta$, and hence in this limit the gapless Goldstone mode should be
asymptotically recovered. We present the calculations of excitation spectrum of $\mathbb{Z}_n$ supersolid below to verify this physical picture.

To evaluate the elementary excitations, we start from the time-dependent GPE,
\begin{equation}
  i\hbar\frac{\partial}{\partial t} \Psi=\left[\hat{H}_{0}+\left(\begin{matrix}
  U_{\uparrow} & 0 \\ 
  0 & U_{\downarrow}
  \end{matrix}\right)\right]\Psi,
  \end{equation} 
 and consider the evolution of perturbation $\delta\Psi$ to the ground state wave function $\Psi$.
  By plugging the perturbed wave function $\Psi'=\Psi+\delta\Psi$
  into the time-dependent GPE and retaining only first-order term of $\delta\Psi$, one will arrive at
  the equation $\delta\Psi$ and $\delta\Psi^*$ satisfy \cite{pethick_Bose_2008}.
   Due to periodic modulation of stripe density, the eigenmodes of perturbation take the following form \cite{li_Superstripes_2013,xia_Metastable_2023}, 
  \begin{align}
  \begin{pmatrix}
  \delta\Psi_\uparrow \\
  \delta\Psi_\downarrow  
  \end{pmatrix} = & e^{-i\mu(k_s)t}\sum_{l=-2L-1}^{2L+1}\begin{pmatrix}
  u_l^\uparrow \\
  u_l^\downarrow
  \end{pmatrix}e^{i (lk_s+q)x}e^{-i\varepsilon(q) t}\nonumber\\
  + & e^{-i\mu(k_s)t} \sum_{l=-2L-1}^{2L+1}\begin{pmatrix}
  v_l^{\uparrow*} \\
  v_l^{\downarrow*}
  \end{pmatrix}e^{i (lk_s-q)x}e^{i\varepsilon(q) t},
  \end{align}
  where $q$ is the Bloch quasimomentum of the excitation. 
  $u_l^{\uparrow\,(\downarrow)}$ and $v_l^{\uparrow\,(\downarrow)}$ are the perturbation amplitudes for spin up (down) component, and 
  the excitation spectrum $\varepsilon(q)$ is numerically obtained by diagonalizing the following Bogoliubov equation 
  that $u_l^{\uparrow\,(\downarrow)}$ and $v_l^{\uparrow\,(\downarrow)}$ satisfy,
\begin{equation}
\mathcal{M}\left( 
\begin{array}{c}
u^{\uparrow } \\ 
u^{\downarrow } \\ 
v^{\uparrow } \\ 
v^{\downarrow }%
\end{array}%
\right) =\varepsilon(q) \left( 
\begin{array}{c}
u^{\uparrow } \\ 
u^{\downarrow } \\ 
v^{\uparrow } \\ 
v^{\downarrow }%
\end{array}%
\right). 
\end{equation}%
Here $u^{\sigma}(x)=\sum_l u_l^\sigma\exp(i(lk_s+q)x)$, and $v^{\sigma}(x)=\sum_l v_l^\sigma\exp(i(-lk_s+q)x)$, with $\sigma=\uparrow, \downarrow$.
The matrix $\mathcal{M}$ reads,
\begin{equation}
\mathcal{M}=\left( 
\begin{array}{cccc}
\mathcal{M}_{11 } & \mathcal{M}_{12} & g\Psi _{\uparrow }^{2} & 
g_{\uparrow \downarrow }\Psi _{\uparrow }\Psi _{\downarrow } \\ 
\mathcal{M}_{12}^{\ast} & \mathcal{M}_{22} & g_{\uparrow \downarrow }\Psi
_{\uparrow }\Psi _{\downarrow } & g\Psi _{\downarrow }^{2} \\ 
-g\Psi _{\uparrow }^{\ast 2} & -g_{\uparrow \downarrow }\Psi _{\uparrow
}^{\ast }\Psi _{\downarrow }^{\ast } & -\mathcal{M}_{11}^{\ast } & \mathcal{M}_{34} \\ 
-g_{\uparrow \downarrow }\Psi _{\uparrow }^{\ast }\Psi _{\downarrow
}^{\ast } & -g\Psi _{\downarrow }^{\ast 2} & \mathcal{M}_{34}^{\ast} & -
\mathcal{M}_{22}^{\ast }%
\end{array}%
\right) , 
\end{equation}%
where $\mathcal{M}_{11}={\hbar^{2}}\left(-i\partial_x-k_{0}\right)^{2}/{2m}+V(x)-\mu +2g|\Psi _{\uparrow }|^{2}+g_{\uparrow \downarrow
}|\Psi _{\downarrow }|^{2}$, $\mathcal{M}_{22}={\hbar^{2}}\left(-i\partial_x+k_{0}\right)^{2}/{2m}+V(x)-\mu +2g|\Psi _{\downarrow }|^{2}+g_{\uparrow \downarrow
}|\Psi _{\uparrow }|^{2}$, $\mathcal{M}_{12}={\Omega}/{2}+g_{\uparrow \downarrow }\Psi_{\uparrow }\Psi _{\downarrow }^{\ast }$, and $\mathcal{M}_{34}=-\mathcal{M}_{12}^{\ast}$.
Note that only the eigenvalues of $\mathcal{M}$ with corresponding eigenvectors satisfy the
normalization condition $\int_0^{d_s}\mathrm{d}x[|u^\uparrow(x)|^2+|u^\downarrow(x)|^2-|v^\uparrow(x)|^2-|v^\downarrow(x)|^2]=1$ represent physical excitations $\varepsilon(q)$.

  \begin{figure}[tbp] \centering
    \includegraphics[width=0.99\linewidth]{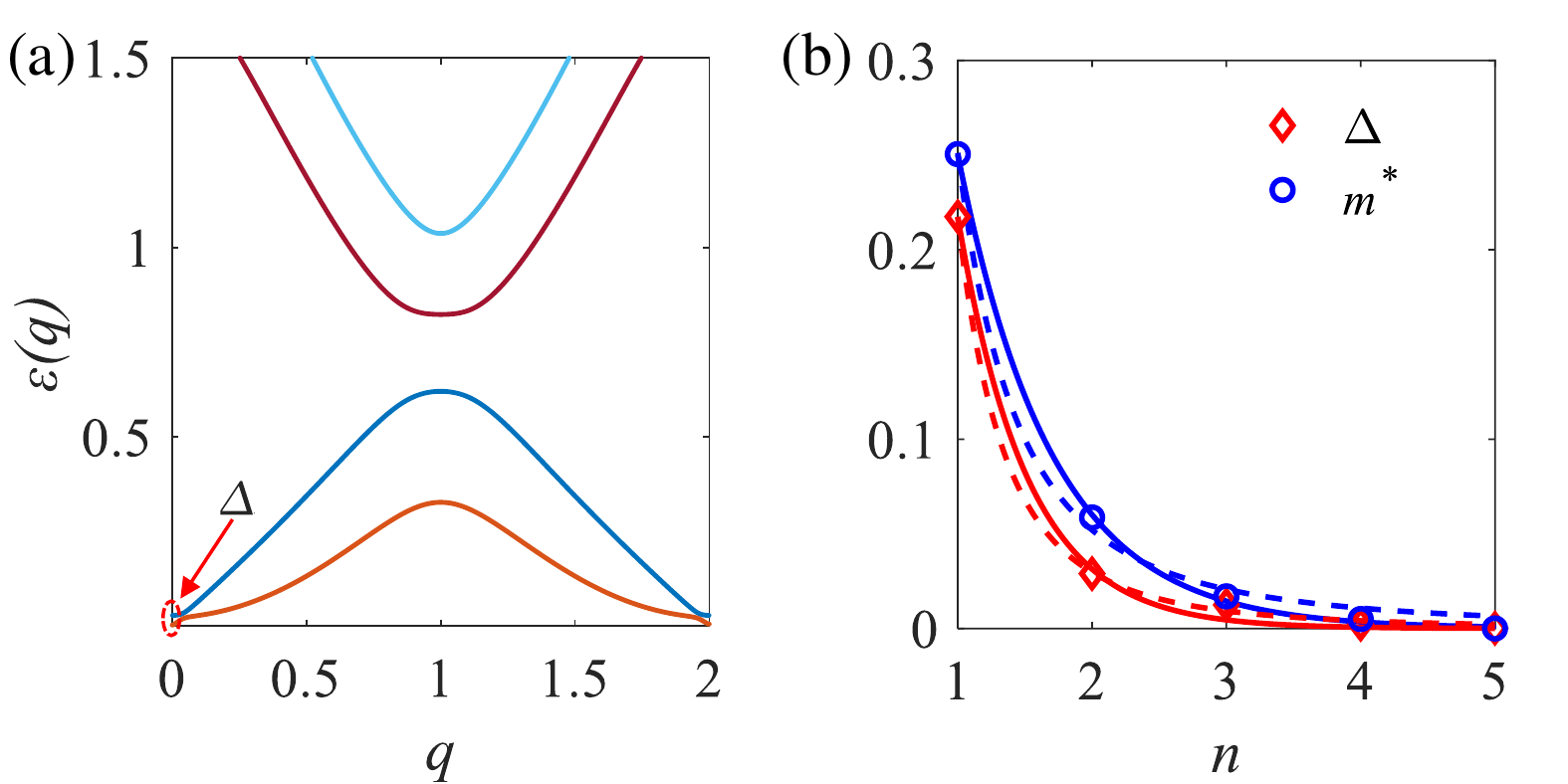}
    \caption{(a) Bogoliubov excitation spectrum for the case of $n=2$. The pseudo-Goldstone gap $\Delta$ is indicated by red arrow. (b) The change of $\Delta$ with respect to $n$ is indicated by red diamonds, fitted with the decay function $\Delta\approx1.504e^{-1.935n}$ ($\Delta\approx0.2175n^{-2.873}$) in red solid (dashed) line. Also shown is the effective mass $m^{*}$ in the effective Hamiltonian by blue circles along with a fitting function $m^{*}\approx1.043e^{-1.427n}$ ($m^*\approx 0.2512n^{-2.272}$) in blue solid (dashed) line. Note that the value of $\Delta$ for large $n$ is less reliable and to improve the accuracy much larger $L$ should be adopted. The other parameters are $V_{0}=2$, $\rho_0g=0.968$, $\rho_0g_{\uparrow\downarrow}=0.88$, and $\Omega=1.294$.}
    \label{fig3}
  \end{figure}

  The excitation spectrum $\varepsilon(q)$ of stripe for $n=2$ is depicted in Fig. \ref{fig3}(a),
 sharing similar feature with the case of $n=1$ in previous work \cite{li_PseudoGoldstone_2021}. 
  In contrast to the gapless Goldstone mode in the case without external potential \cite{li_Superstripes_2013}, 
  the lattice potential breaks the continuous translation symmetry of the system and the excitation spectrum features one branch of gapless excitation and one branch of
  gapped excitation. The gapless branch is originated from the spontaneous breaking of
  U(1) gauge symmetry still present here, being a common feature of superfluid; the gapped branch is due to the explicit breaking of spatial translation symmetry by external potential.
In comparison with the $n=1$ case, the major difference here is the change of pseudo-Goldstone gap $\Delta$ in long wavelength for different $n$.
  The dependence of $\Delta$ on $n$ is shown in Fig. \ref{fig3}(b) (red diamonds),
  where the $\Delta$ dependence on $n$ can be approximately fitted by a fast-decaying function (both the exponential fitting and power-law fitting 
  are good within the finite set of data). In fact, for $n\geq3$, the gap becomes negligibly small.
  So for large $n$, the pseudo-Goldstone mode resembles the gapless Goldstone mode, consistent with the expectation that $\mathbb{Z}_n$ group
  becomes a better approximation for U(1) group in large $n$ limit.
  
  The decay behaviour of $\Delta$ with $n$ can be qualitatively understood within the framework of effective Hamiltonian which describes the low-energy physics of this $\mathbb{Z}_n$ symmetry broken supersolid. We first consider the small deviation of $\theta$ from one of the
  local effective potential minima in Fig. \ref{fig2}, accompanied by the increase of effective potential energy due to the $V(x)$ term. Then, taking also into account the spatial variation of the
  effective field $\theta$ and neglecting other irrelevant dynamical variables \cite{jian_Paired_2011}, 
  the low-energy effective Hamiltonian becomes 
  \begin{equation}
  \mathcal{H}_\mathrm{eff}\sim\frac{\rho_0}{m}\sum_{l=-2L-1}^{2L+1} l^2 a_l^2 \left(\partial_x\theta\right)^2+E_v(\theta).
  \end{equation}
 Here the first term accounts for the energy associated with the spatial variation of $\theta$,
 obtained by plugging the wave function in Eq. \ref{eq5} into the kinetic part of $\hat{H}_0$ (the first two terms in Eq. \ref{eq1}), and the second term is the effective potential $E_v(\theta)\approx E_v(\theta_0)+2\rho_0 n^2 V_{0}\sum_{l_{1}-l_{2}=2n}^{}a_{l_{1}}a_{l_{2}}(\delta\theta)^2$ around one potential minimum $\theta_0$. By recasting the above effective Hamiltonian into the generic form which describes a massive field, i.e., $\mathcal{H}_\mathrm{eff}\sim \rho_0\left[\frac{1}{m}\sum_l l^2 a_l^2\left(\partial_x\theta\right)^2+m^*(\delta\theta)^2\right]$, one can define the effective mass $m^*$ or the pseudo-Goldstone gap as 
    \begin{equation}
  m^*=2n^2 V_{0}\sum_{l_{1}-l_{2}=2n}^{}a_{l_{1}}a_{l_{2}}.
  \end{equation}
So, through coupling with external periodic potential, the $\theta$ field changes from massless to massive \cite{jian_Paired_2011}, with the mass tunable by both potential strength $V_0$ and integer $n$.
 With the above expression, one can infer that the dependence of $m^*$ on $n$ is originated from the dependence of coefficient $a_l$ on $l$.
 Indeed, both our numerical calculations and perturbative analysis \cite{martone_Supersolid_2021} show that the coefficient $a_l$ decays almost
   exponentially with $l$. With this information, one can demonstrate that the gap $m^*$ decays fast with the increase of $n$. Both the value of $m^*$ and the decay behaviour are in qualitative agreement with $\Delta$, as shown in Fig. \ref{fig3}(b). The effective Hamiltonian method hence provides a physical
   picture in understanding the origin of decay behaviour in the pseudo-Goldstone gap.

\section{Transition between degenerate ground states}
\label{sec4}

\begin{figure}[tbp] \centering
	\includegraphics[width=0.99\linewidth]{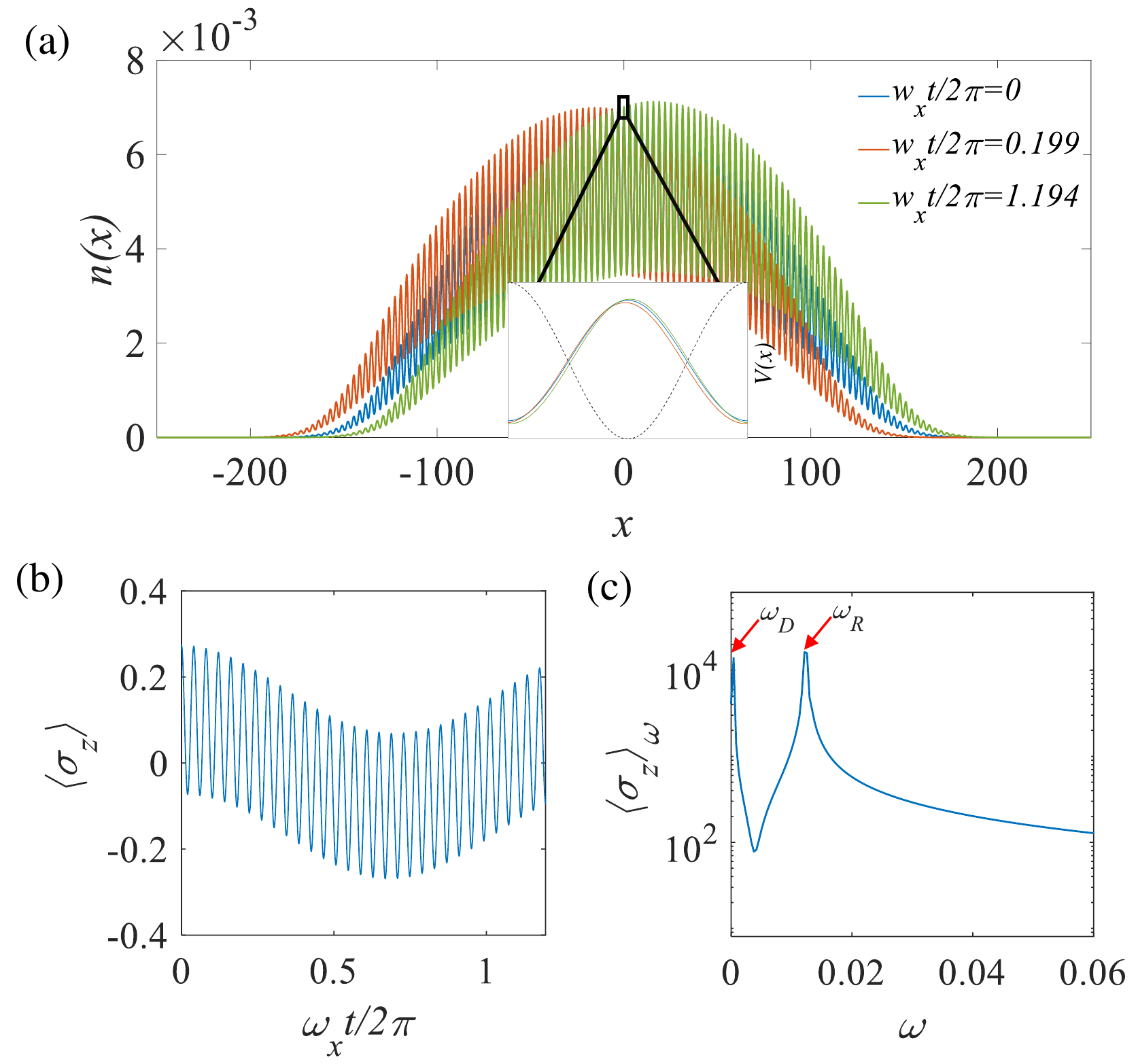}
	\caption{(a) Stripe density distribution $n(x)$ at three different moments indicated by the legend in the case of $n=1$. 
  (b) Temporal evolution of $\left\langle \sigma_{z}\right\rangle $. 
  (c) Frequency spectrum of $\left\langle \sigma_{z}\right\rangle $ in (b) obtained by Fourier transform. 
  The near-zero frequency $\omega_{\mathrm{D}}=0.00042$ (corresponding to dipole mode) and non-zero frequency $\omega_{\mathrm{R}}=0.012$ (corresponding to Rabi-like oscillation) are indicated by red arrows. 
  The other parameters are $\rho_0g=0.968$, $\rho_0g_{\uparrow\downarrow}=0.88$, $\Omega=1.294$, 
  $V_{0}=-0.08$, $\lambda=0.005$, and $\omega_{x}=0.0005$.}
	\label{fig4}
\end{figure}

In the last part of this paper, we focus on the experimental signature of $\mathbb{Z}_n$ supersolid. 
In realistic experiments, there is usually a harmonic trap which confines the supersolid, and
the pseudo-Goldstone mode excitation will consequently be modified. Recently, it has been shown that,
by applying a spin-dependent perturbation similar to a Zeeman energy, 
the Goldstone mode in the absence of periodic potential can be excited and directly revealed from the translation of the stripes \cite{geier_Exciting_2021}. 
Here, in the presence of both harmonic trap and periodic potential, 
we propose that the pseudo-Goldstone mode can also be excited by adding the same spin perturbation. 
In particular, the dynamical behaviour of $\mathbb{Z}_n$ supersolid has distinct features for $n=1$ and $n\geq2$, which provide an experimental
signature for their observation.

To be specific,  
we first find the ground state of stripes with the external potential $V(x)=m\omega_{x}x^{2}/2+V_{0}\cos(2nk_{s}x)$ 
and the perturbation $-\lambda E_{r}\sigma_{z}$, using the method of imaginary time
evolution.
Then, at moment $t=0$, the perturbation is suddenly released, and to obtain the real-time evolution of stripes, 
we numerically solve the following time-dependent GPE
\begin{equation}
  i\hbar\frac{\partial}{\partial t} \Psi=\left[\hat{H}_{0}+\left(\begin{matrix}
  U_{\uparrow} & 0 \\ 
  0 & U_{\downarrow}
  \end{matrix}\right)+\frac{1}{2}m\omega_{x}x^{2}\right]\Psi.
\end{equation}
This procedure amounts to preparing the system in a weakly perturbed initial state above ground state, and then monitoring how the weak excitation evolves in time. The focus is on the temporal evolution of two experimental observables, i.e.,
the overall density
$n(x)=|\Psi_\uparrow|^{2}+|\Psi_\downarrow|^{2}$ and the expectation value of $\sigma_{z}$ $\left\langle \sigma_{z}\right\rangle  = \int\mathrm{d}x\left(\Psi^{\dagger} \sigma_{z} \Psi\right)$.

\begin{figure}[tbp] \centering
	\includegraphics[width=0.9\linewidth]{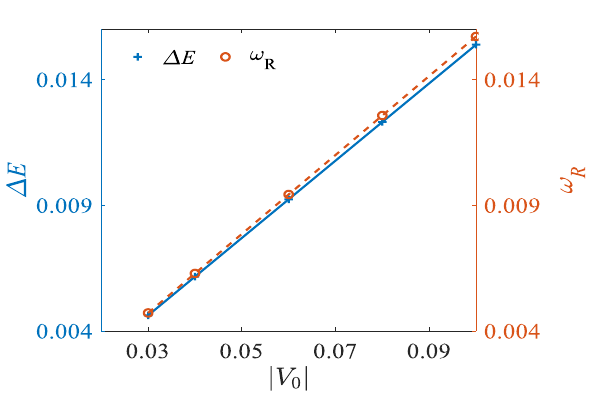}
	\caption{The change of energy gap $\Delta E$ at dispersion minima of spin-orbit-coupled Bloch band in Fig. \ref{fig1}(b) and 
  non-zero oscillation frequency $\omega_{\mathrm{R}}$ in Fig. \ref{fig4}(c) are denoted by blue solid line (with ``+'' symbols) and orange dashed line (with circles), respectively, in the case of $n=1$.
 The other parameters are $\rho_0g=0.968$, $\rho_0g_{\uparrow\downarrow}=0.88$, and $\Omega=1.294$.}
	\label{fig5}
\end{figure}

For the case of $n=1$, we depict in Fig. \ref{fig4}(a) the overall density $n(x)$ at three different moments which reveal 
the very weak excitation of the translational motion of the stripes,
in stark to the case without periodic potential \cite{geier_Exciting_2021}.
In that case, the stripe density displays an obvious translational motion at constant velocity caused by the spin-dependent perturbation.
 Here, the spatial shift of stripe density is strongly suppressed. This can be understood from the
naive picture that the spatial shift of stripe density is associated with the deviation of $\theta$ from the minimum of effective potential energy in 
Fig. \ref{fig2}(a2), which encounters a large potential barrier and hence is suppressed. For weak perturbation, only small oscillation of $\theta$ around the
potential minimum is excited and the spatial shift of stripe density is nearly invisible.

\begin{figure}[tbp] \centering
	\includegraphics[width=0.99\linewidth]{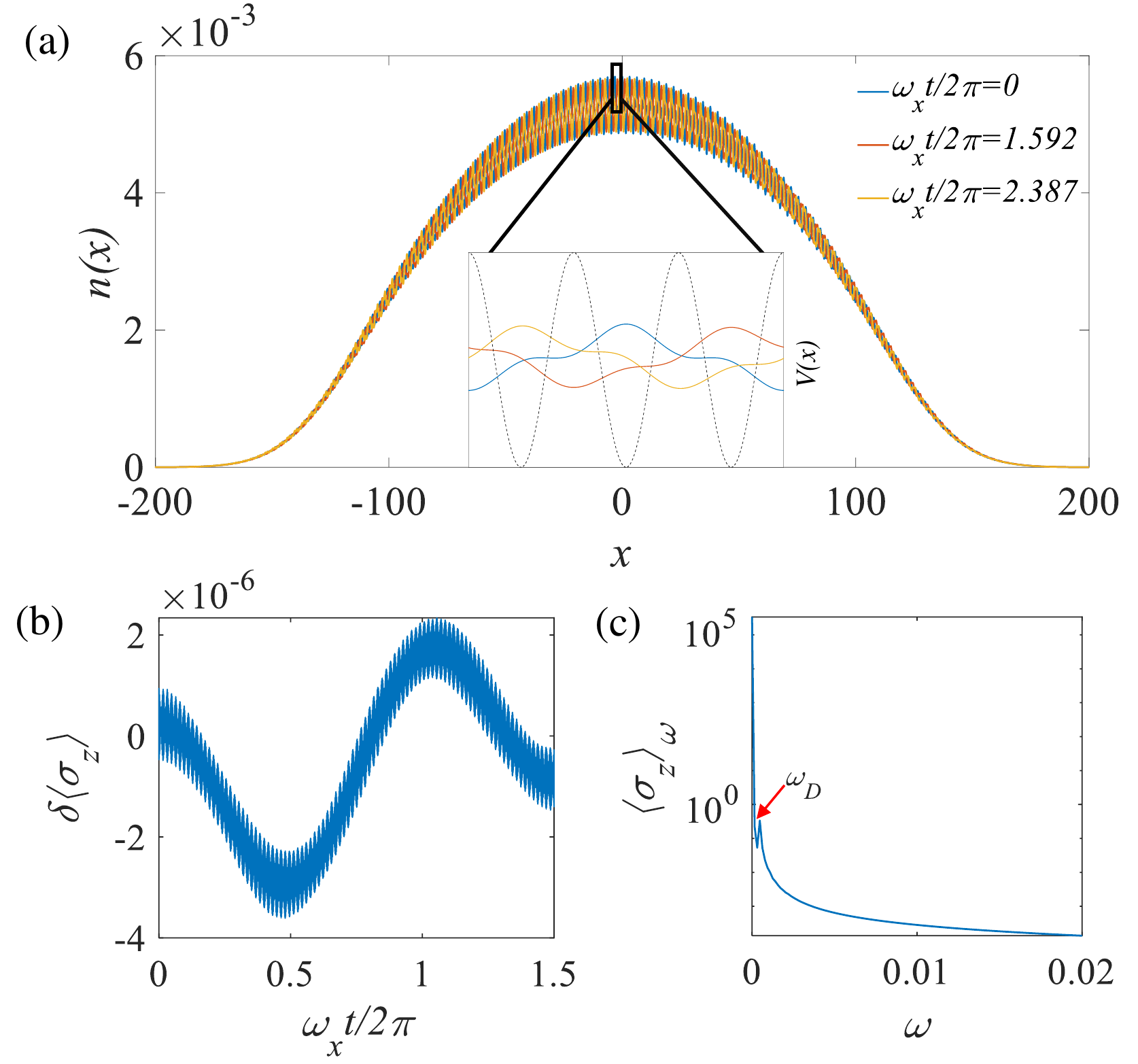}
	\caption{(a) Stripe density distribution $n(x)$ at three different moments denoted by the legend showing the spatial shift of stripe density. 
  (b) Temporal evolution of $\delta\left\langle \sigma_{z}\right\rangle=\left\langle \sigma_{z}\right\rangle-\left\langle \sigma_{z}\right\rangle_0$,
  with $\left\langle \sigma_{z}\right\rangle_0\approx0.84$ subtracted for better illustration.
  (c) The frequency spectrum of $\left\langle \sigma_{z}\right\rangle$ in the case of $n=3$. 
  The near-zero frequency $\omega_{\mathrm{D}}=0.00047$ (corresponding to dipole mode) is indicated by red arrow. The other parameters are
  $\rho_0g=0.968$, $\rho_0g_{\uparrow\downarrow}=0.88$, $\Omega=0.4$, $V_{0}=-0.3$, $\lambda=0.0003$, and $\omega_{x}=0.0005$.}
	\label{fig6}
\end{figure}

In Figs. \ref{fig4}(b) and (c), we also show
the temporal oscillation of $\left\langle \sigma_{z}\right\rangle$ and corresponding frequency spectrum obtained by 
Fourier transform of $\left\langle \sigma_{z}\right\rangle_t$. 
The frequency spectrum $\left\langle \sigma_{z}\right\rangle_{\omega}$ shows two dominant non-zero peaks, 
one corresponding to a near-zero frequency and the other to a non-zero frequency (corresponding to the left and right peaks in
Fig. \ref{fig4}(c), respectively). 
By comparing with the case without periodic potential \cite{geier_Exciting_2021}, one can infer that the near-zero frequency corresponds to
the dipole mode in the presence of harmonic trap, with frequency $\omega_\mathrm{D}\approx0.8\omega_x$ for the specific parameters here. Interestingly, there is also another non-zero frequency
$\omega_{\mathrm{R}}$,
which is absent in the case without periodic potential.
To shed light on the nature of this frequency, in Fig. \ref{fig5}, 
we compare the value of this frequency and the energy gap $\Delta E$ at the BZ boundary of spin-orbit-coupled Bloch band in Fig. \ref{fig1}(b).
Remarkably, these two quantities are very close, reflecting the fact that 
the fast oscillation of $\left\langle \sigma_{z}\right\rangle_t$ results from the Rabi-like oscillation between the single-particle ground state and 
excited state by weak perturbation, similar to the phenomenon observed in Ref. \cite{bersano_Experimental_2019}. 
The combination of $\omega_\mathrm{D}$ and $\omega_{\mathrm{R}}$ leads to a beating effect of $\left\langle \sigma_{z}\right\rangle $ as shown in Fig. \ref{fig4}(b).

Next, for comparison, we turn to discuss the $n\neq1$ case (see Fig. \ref{fig6} for $n=3$). 
As mentioned in Sec. \ref{sec2}, the ground state is degenerate in this case. 
Interestingly, the time evolution of $n(x)$ also explicitly reveals the transition between different degenerate states which resembles the spatial shift of stripe in the case
of vanishing periodic potential \cite{geier_Exciting_2021}. In terms of the dynamical variable $\theta$, this corresponds to 
drifting of $\theta$ from one potential minimum to another by overcoming the potential barrier between degenerate minima, as illustrated in Fig. \ref{fig2}(c2).
Note that for the case of $n=3$, the effective potential barrier is rather small (see Fig. \ref{fig2}(c2)), and thus a very weak excitation is enough for the
transition between degenerate states.
As shown in Figs. \ref{fig6}(b) and (c), 
other features are also similar to the case of vanishing periodic potential, such as
the locking of polarization over a long time and a near-zero frequency corresponding to
the dipole mode \cite{geier_Exciting_2021}. It is noted that the amplitude of dipole mode frequency is very small, 
because $\left\langle \sigma_{z}\right\rangle $ is nearly constant during time evolution.
In addition, the larger frequency $\omega_{\mathrm{R}}$ in the case of $n=1$ is absent here, which confirms our speculation that $\omega_{\mathrm{R}}$
corresponds to the gap brought by external potential at dispersion minima in Fig. \ref{fig1}(b), and also serves as a clear-cut signature to
distinguish between cases of $n=1$ and $n\ge2$. 
In conclusion, the overall feature of stripe dynamics under weak excitation is similar to the case
without periodic potential, which agrees with the expectation that $\mathbb{Z}_n$ symmetry asymptotically approaches U(1) symmetry and the fast
decay of excitation gap in Fig. \ref{fig3}(b).

\section{Conclusion and Acknowledgement}
\label{sec5}

In summary, we have proposed the concept of $\mathbb{Z}_n$ symmetry broken supersolid in the stripe phase of spin-orbit-coupled BEC under external
periodic potential, and explored in depth the intriguing properties of this $\mathbb{Z}_n$ supersolid, in comparison with conventional supersolid
resulting from continuous translation symmetry breaking.
We have explicitly demonstrated the existence of degenerate ground states in the presence of periodic potential whose period is $1/n$ of intrinsic stripe period.
These degenerate ground states differ by spatial translation of stripe density over the potential period and can be understood in terms
of discrete values for the relative phase in the stripe wave function, corresponding to multiple minima of effective potential.
Afterwords, we show that the excitation gap of pseudo-Goldstone mode decays fast with $n$ and explain this behaviour based on the low-energy effective
Hamiltonian. We further consider the effect of harmonic trapping potential and study the dynamical features of the $\mathbb{Z}_n$ supersolid under a spin-dependent perturbation.
The spin polarization oscillations feature two dominant frequencies in the case of $n=1$,
and have only one dominant frequency in the cases of $n\ge2$.
We also numerically observe the transition between the degenerate ground states
under a weak excitation in the cases of $n\ge2$. With the successful experimental realization of supersolid in a spin-orbit-coupled BEC and the technique of imposing subwavelength optical
lattice, our prediction can be readily tested in current experiments.
In addition, this model also provides an
experimentally feasible system to simulate the physical consequence associated with a generic $\mathbb{Z}_n$ symmetry breaking, which may help elucidate similar phenomena in other branches of physics, e.g., particle physics and cosmology.

This work is supported by the National Key Research and 
Development Program of China (Grant No. 2022YFA1405304), the National Natural
Science Foundation of China (Grant No. 12004118), and the Guangdong Basic and
Applied Basic Research Foundation (Grants No. 2020A1515110228 and No. 2021A1515010212).

\end{document}